\documentclass[11pt,preprint]{aastex}

\usepackage{graphicx}
\usepackage{xcolor}
\usepackage{natbib}

\def\apj{\rm ApJ}
\def\apjl{\rm ApJL}

\def\aap{\rm AAP}
\def\aj{\rm AJ}
\def\mnras{\rm MNRAS}
\def\nat{\rm Nature}
\def\pasj{\rm PASJ}

\def\physrep{\rm Phys. Rep}

\shorttitle{X-ray Temperature TDE}
\shortauthors{Dai et al.}

\begin{document}

\title{Soft X-ray Temperature Tidal Disruption Events \\
from Stars on Deep Plunging Orbits}

\author{Lixin Dai}
\affil{Department of Physics and Joint Space-Science Institute, \\
University of Maryland, College Park, MD 20742; \\ cosimo@umd.edu}

\author{Jonathan C. McKinney}
\affil{Department of Physics and Joint Space-Science Institute, \\
University of Maryland, College Park, MD 20742}

\author{M. Coleman Miller}
\affil{Department of Astronomy and Joint Space-Science Institute, \\
University of Maryland, College Park, MD 20742}

\begin{abstract}

One of the puzzles associated with tidal disruption event candidates (TDEs) is that there is a dichotomy between the color temperatures of ${\rm few}\times 10^4$~K for TDEs discovered with optical and UV telescopes, and the color temperatures of ${\rm few}\times 10^5 - 10^6$~K for TDEs discovered with X-ray satellites.  Here we propose that high-temperature TDEs are produced when the tidal debris of a disrupted star self-intersects relatively close to the supermassive black hole, in contrast to the more distant self-intersection that leads to lower color temperatures. In particular, we note from simple ballistic considerations that greater apsidal precession in an orbit is the key to closer self-intersection. Thus larger values of $\beta$, the ratio of the tidal radius to the pericenter distance of the initial orbit, are more likely to lead to higher temperatures of more compact disks which are super-Eddington and geometrically and optically thick.  For a given star and $\beta$, apsidal precession also increases for larger black hole masses, but larger black hole masses imply a lower temperature at the Eddington luminosity.  Thus the expected dependence of the temperature on the mass of the black hole is non-monotonic.  We find that in order to produce a soft X-ray temperature TDE, a deep plunging stellar orbit with $\beta> 3$ is needed and a black hole mass of $\lesssim 5\times 10^6 M_\odot$ is favored.  Although observations of TDEs are comparatively scarce and are likely dominated by selection effects, it is encouraging that both expectations are consistent with current data.
\end{abstract}


\pagebreak

\section{Introduction}

The discovery of roughly two dozen TDEs with X-ray and optical/UV telescopes (e.g., \citealt{1996A&A...309L..35B, 1999A&A...349L..45K,2008A&A...489..543E, 2011ApJ...741...73V, 2009ApJ...698.1367G, 2012Natur.485..217G, 2012MNRAS.420.2684C, 2012A&A...541A.106S, 2013MNRAS.435.1904M, 2014ApJ...780...44C, 2014ApJ...793...38A, 2014MNRAS.445.3263H, 2015arXiv150701598H}) has afforded us the opportunity to study accretion over a wide range of rates and is also promising for the discovery and characterization of otherwise quiescent supermassive black holes (SMBHs) \citep{1997ApJ...489..573L}. Standard treatments predict a temperature that is ${\rm few}\times 10^5$~K near the peak of the burst and decreases with the luminosity \citep{1988Natur.333..523R, 1989IAUS..136..543P, 1990ApJ...351...38C, 1999ApJ...514..180U}. However, although X-ray detected TDEs reach this temperature, the TDEs discovered via optical/UV observations have lower temperatures ${\rm few}\times 10^4$~K, and those temperatures remain nearly steady even as the luminosity of the sources drop by $1-2$ orders of magnitude \citep[e.g.,][]{2012Natur.485..217G, 2014ApJ...793...38A}. This has been explained as a consequence of either an optically thick shroud of gas at many times the radius of the disk \citep{1997ApJ...489..573L} or a wind from the disk \citep{2009MNRAS.400.2070S, 2015ApJ...805...83M, 2015arXiv150603453M}.  Why do the processes which reduce the temperature in optical/UV detected TDEs fail to operate for the X-ray detected TDEs that have temperatures of ${\rm few}\times 10^5-10^6$~K?

The ranges in length and time scales required for full hydrodynamic simulations of debris circularization in tidal disruptions of a main-sequence star by a SMBH from a marginally bound orbit mean that such simulations are currently computationally infeasible. The main reason is that such simulations need to be able to follow the debris orbit with radius $r \sim 100-1000 R_g$, and at the same time need to resolve the thickness of the stream which, if is self-gravitating, is much less than $r$ \citep{1994ApJ...422..508K, 2014ApJ...783...23G}. The addition of realistic cooling mechanisms and general relativistic hydrodynamics would make simulations even more time-consuming. However, studies of stars disrupted by SMBHs from initially bound orbits \citep{2015arXiv150104635B, 2015arXiv150105207H} and white dwarfs disrupted by intermediate mass black holes \citep{2009ApJ...697L..77R, 2015ApJ...804...85S} have revealed that debris circularization is more difficult than previously thought. The nozzle shock and the instabilities at pericenter are not strong enough to circularize the debris quickly \citep{1994ApJ...422..508K, 2014ApJ...783...23G, 2015ApJ...804...85S}. Instead, tidal stream intersection is the most effective way to produce shocks and dissipate the debris orbital energy \citep{2015ApJ...804...85S, 2015arXiv150104635B,2015arXiv150105207H}.  This intersection is a result of the orbital apsidal precession of the debris on the same plane around the SMBH. However, if apsidal precession is small and intersection happens far from the pericenter of the initial orbit, then the collision is mild and circularization can be delayed substantially (see, e.g., \citealt{2015ApJ...804...85S} and the Newtonian simulation of \citealt{2015arXiv150104635B}). In this case, most of the matter would take much longer to accrete onto the black hole than it would in the standard picture. As a consequence, the luminosity and the disk temperature would both be less than those seen in X-ray detected TDEs. 

Here we propose that the high temperatures of some TDEs can be explained in a picture in which the debris disk is small due to relatively large apsidal precession.  In particular, we argue that if the disk is small enough that the initial temperature exceeds $\sim 10^5$~K, then (unlike when $T={\rm few}\times 10^4$~K) the opacity is only weakly sensitive to the temperature, which eliminates the strong dependence of wind rates on temperatures that features in one explanation for the lower temperatures seen in optical/UV TDEs \citep{2015ApJ...805...83M}.  In Section 2 we perform calculations of ballistic motion to explore how debris stream intersection and the consequent disk size depend on the black hole mass $M$ and the stellar orbit penetration parameter $\beta$. In Section 3 we calculate the disk temperature, where we show that high-$\beta$ encounters, particularly with black holes of $M \sim 10^6~M_\odot$, are the best candidates for TDEs with temperatures $\gtrsim {\rm few}\times 10^5$~K. Further discussions and some observational considerations are given in Section 4.

\section{Tidal Stream Intersection: First Order Calculations}

\begin{figure}
\centering
 \includegraphics[width=6in]{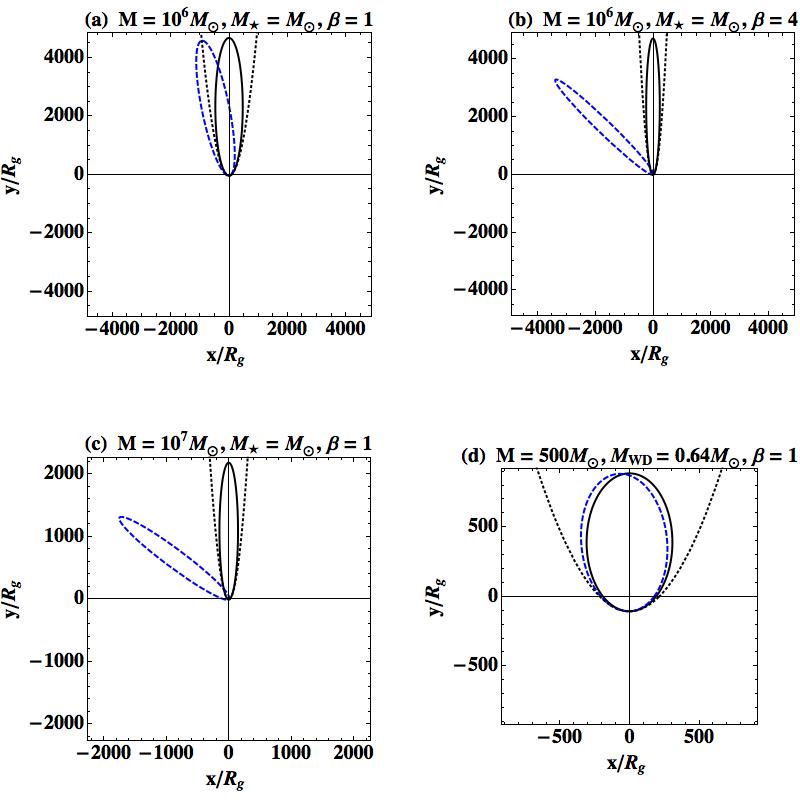}
\caption{The intersection of tidal streams for different setups. The black hole is at the origin. The black solid ellipse is the most bound orbit in a Newtonian potential. The black dotted curve is the parabolic trajectory of the center of the star. The blue dashed ellipse is the ballistic trajectory of the most bound orbit with a periastron precession of angle $\phi$ (Equation (1)). Panels (a), (b), and (c) represent the scenarios for a SMBH and a solar type star. Panel (d) represents an IMBH-white dwarf disruption scenario as in \citet{2015ApJ...804...85S}.  Stream-stream intersection occurs where the black solid line and blue dashed line intersect. In a deep plunge (panel (b)) or when the black hole is more massive (panel (c)) the stream intersects closer to the hole.}
\label{Intersection}
\end{figure}

As the nozzle shock is weak, the trajectory of the debris is nearly ballistic until the tidal stream self-intersects. Although some self-intersection would occur even in Newtonian gravity because there is a spread in debris binding energy, the dynamics of the intersection are dominated in our case by general relativistic pericenter precession.  If we assume for simplicity a non-rotating (Schwarzschild) black hole (rotational corrections are mild unless the pericenter is very close to the hole), then the precession angle $\phi$ over a single orbit is, to first order,
\begin{equation}
\phi = 6 \pi M / (a (1-e^2)),
\end{equation}
(\citealt{1973grav.book.....M}), where $a$ is the semi-major axis of the orbit, and $e$ is the orbital eccentricity.  Here and henceforth we use geometrized units in which $G=c=1$. If the orbit is close to the hole, the exact precession rate is larger than the rate given by this expression, but the differences are mild for $a(1-e)>10M$. In reality the ellipse precesses continuously, but for the high-eccentricity orbits we consider almost all the precession occurs at pericenter. We therefore treat the debris orbit as a closed ellipse with an instantaneous shift of $\phi$ at pericenter passage, as illustrated in Figure~\ref{Intersection}. 

The head of the stream essentially follows the most bound debris orbit, and after passing pericenter it intersects the trailing part of the incoming stream (the shifted orbit is represented by the blue ellipse in Figure~\ref{Intersection}). The incoming stream lies between the most bound debris orbit (the black ellipse) and the marginally bound debris orbit (the black dotted curve). If the initial trajectory of the star is nearly parabolic, intersection typically happens in a time after disruption of $1-1.5 P_{\rm mb}$ \citep{2015ApJ...804...85S,2015arXiv150105207H,2015arXiv150104635B,2015ApJ...809..166G}, where $P_{\rm mb}$ is the orbital period of the most bound orbit. 

We now study the dynamics of the most bound orbit for the disruption of a star of mass $m_*=10^0~m_{*,0}M_\odot$ and a radius $R_*=10^0~R_{*,0}R_\odot$, where $R_\odot$ is the radius of the Sun, by a SMBH with mass $M = 10^6~M_6 M_\odot$.

\subsection{The most bound orbit}
A star on an initially parabolic orbit has zero specific binding energy:
\begin{equation}
E = 1/2 \ v^2_T- M/R_T = 0,
\end{equation}
where $R_T\approx R_*(M/m_*)^{1/3}$ is the tidal radius and $v_T$ is the orbital speed at $R_T$.  If we neglect the rotation of the star and the redistribution of energy during the compression / rebound processes, the most bound orbit has energy:
\begin{equation}
E_{\rm mb} = 1/2 \ v^2_T- M/(R_T - R_*) \approx -MR_*/R^2_{\it T},
\end{equation}
because $R_{\it T} \gg R_\star $.  The semi-major axis and eccentricity of this orbit are 
\begin{equation}
a_{\rm mb} = R^2_T/2R_* \approx 3.5\times 10^{14} ~{\rm cm} ~R_{*,0}M_6^{2/3}m_{*,0}^{-2/3}
\end{equation}
and
\begin{equation}
e_{\rm mb} = 1- R_p/ a_{\rm mb} \approx 1 - 2/\beta \times (m_*/M)^{1/3} = 1- 0.02 m^{1/3}_{*,0}M^{-1/3}_6 \beta^{-1}.
\end{equation}
Here $R_p$ is the pericenter distance, and we have defined the penetration factor $\beta\equiv R_T/R_p$. The orbital period of the most bound debris is:
\begin{equation}
P_{\rm mb} = 2 \pi \sqrt{a^3_{\rm mb}/M}=0.11~{\rm yr}~R_{*,0}^{3/2}M_6^{1/2}m_{*,0}^{-1}
\end{equation}
(note that we neglect the weak increase in $P_{\rm mb}$ with increasing $\beta$ found by \citealt{2013ApJ...767...25G}). Therefore, the most bound orbit in a main-sequence star--SMBH disruption is highly eccentric and its orbital period ranges between $\sim 1$ month ($M_6=1$) and $\sim 1$ year ($M_6=100$). 

\subsection{The intersection radius and collision angle}

\begin{figure}
\centering
 \includegraphics[width=6in]{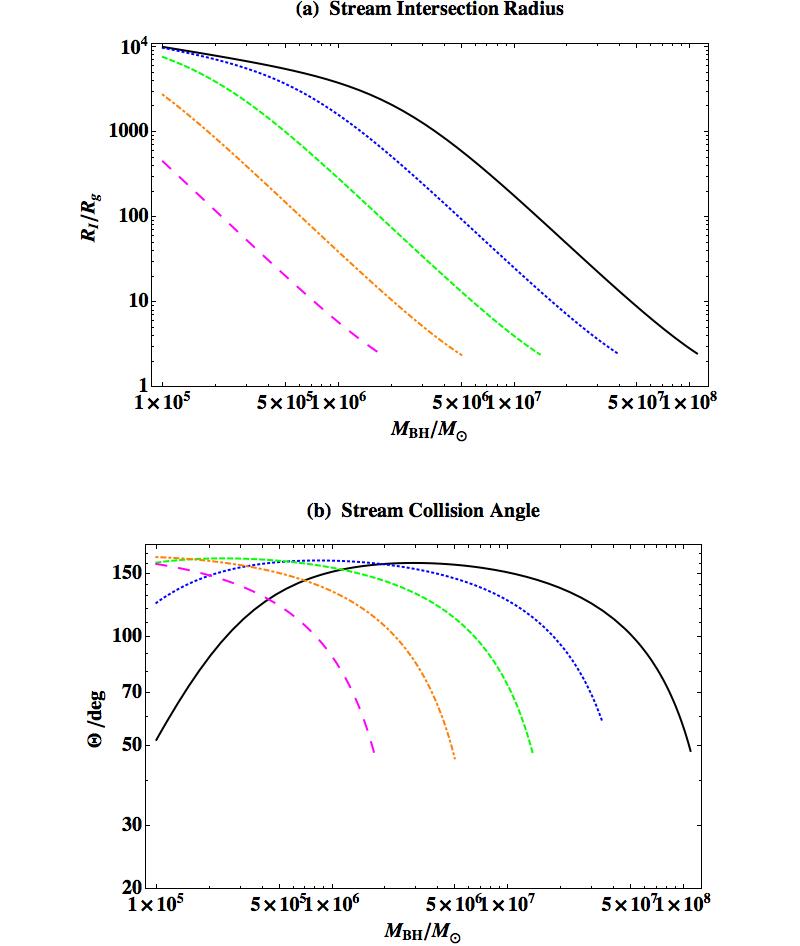}
\caption{The intersection radius and collision angle of the most bound orbit with itself after apsidal precession. Different colors and line styles represent different $\beta$: 1 - black solid, 2 - blue dotted, 4 - green dashed, 8 - orange dot-dashed , and 16 - magenta long-dashed. Panel (a) shows that the interaction radius decreases with larger $\beta$. Panel (b) shows that the collision angle in most cases is between 40 and 160$^\circ$. The energy loss is more efficient when the collision angle is large.}
\label{RadiusAngle}
\end{figure}

As discussed previously, the narrowness of the tidal stream means that the intersection radius of the tidal streams is given by apsidal precession. Treating the most bound orbit as an ellipse with an instantaneous pericenter shift $\phi$, a geometrical calculation shows that the original ellipse and the shifted ellipse intersect at a radius
\begin{equation}
R_I = \frac{(1+ e_{\rm mb}) R_T}{\beta (1- e_{\rm mb} \cos(\phi/2))}.
\end{equation}
Figure~\ref{RadiusAngle}(a) shows that this stream intersection radius changes sensitively with $M$ and $\beta$ (hereafter we use a solar-type star in calculations).  For more massive holes, the tidal radius relative to the size of the hole decreases, so stronger apsidal precession reduces the intersection radius in units of the black hole gravitational radius $R_g\equiv GM/c^2$. For higher $\beta$, $R_p$ is smaller, which again implies stronger precession and closer intersection.

The intersection angle $\Theta$ of the outgoing most bound orbit with the incoming stream is given by
\begin{equation}
\label{collisionangle}
\cos\Theta = \frac{1-2 \cos(\phi/2) e_{\rm mb} + \cos\phi \ e^2_{\rm mb}} {1-2 \cos(\phi/2) e_{\rm mb} + e^2_{\rm mb}}.
\end{equation}
$\Theta$ is plotted as a function of $M$ and $\beta$ in Figure~\ref{RadiusAngle}(b). Streams usually collide with $\Theta \sim 40-160^\circ$ if $R_p\gg R_g$. The larger the intersection angle, the more effectively the collision reduces the orbital energy of the debris.

\subsection{Energy dissipation at stream intersection}

Some fraction of orbital energy is lost during stream collision due to shocks and instabilities. The exact fraction depends on the collision angle and velocity/density contrast of the streams \citep{1994ApJ...422..508K}. Here we adopt an inelastic collision model in which the outgoing and incoming streams have similar mass \citep[cf.][]{2015ApJ...804...85S}. From momentum conservation, the post-collision speed of the stream at the intersection point is
\begin{equation}
v_f = v_i \cos(\Theta/2)\; ,
\end{equation}
where the speed $v_i$ of the streams just before collision is given by
\begin{equation}
- \frac{GM} {2a_{\rm mb}}=  - \frac{GM} {R_I} + \frac{1}{2} v_i^2.
\end{equation}

\begin{figure}
\centering
 \includegraphics[width=6in]{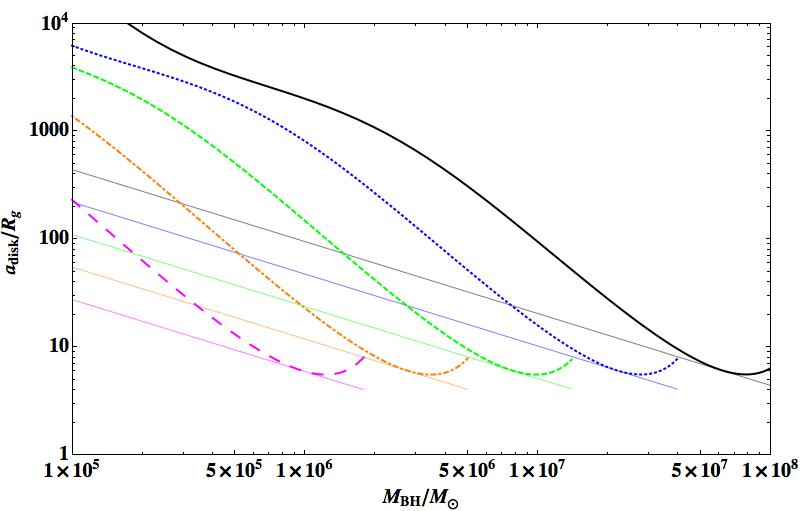}
\caption{The characteristic sizes of the elliptical disk after stream intersection are represented by the thick curves. The color and line style scheme are the same as in the previous figure. For comparison, the thin lines show the corresponding classical circularization radius for parabolic TDEs. This figure shows that the disk size greatly decreases with increasing $\beta$.}
\label{Disk}
\end{figure}

The specific energy loss in the collision is $\Delta E = \frac{1}{2} v_i^2 \sin^2(\Theta/2)$. The debris forms an elliptical disk after collision, and the semi-major axis of this elliptical disk is
\begin{equation}
a_{\rm disk} = \frac {R_I}{2 \sin^2(\Theta/2)} \frac{1}{1+  \frac {R_I}{2 a_{\rm mb}} \cot^2(\Theta/2) }.
\end{equation}
For deep encounters, $a_{\rm mb} \gg R_I$ and thus
\begin{equation}
a_{\rm disk} \approx \frac {R_I}{2 \sin^2(\Theta/2)}.
\end{equation}
We plot this characteristic size of debris disk in Figure~\ref{Disk}. The disk size shrinks rapidly as $M$ or $\beta$ increases. This is because the energy loss is large when the intersection is close and the collision speed is fast. When $\beta$ is large, the disk radius is comparable to the classical circularization radius $R_c = 2 / \beta \times R_T$, for which the debris materials are fully circularized.

\section{Debris Disk Accretion and Temperature}

The evolution of the post-collision elliptical disk depends on further stream--stream and stream--disk interactions. The circularization timescale has traditionally been assumed to be several dynamical timescales of the most bound orbit: 
\begin{equation}
T_{\rm circ} = n \times P_{\rm mb},
\end{equation}
with $n = 2 -10$ \citep{1989ApJ...346L..13E,1999ApJ...514..180U}. However, \citet{2015ApJ...804...85S} found that for a white dwarf disrupted by an intermediate-mass black hole the disk is only partially circularized after $n>10$ if the stream collision is mild. Though it is unclear how this result extrapolates to a higher mass ratio TDE, it is plausible that the circularization timescale will be shorter for higher $\beta$ where the collision is stronger.  For our calculation we assume at $T_{\rm circ} = 5 P_{\rm mb}$ the debris materials are largely circularized and the disk becomes stable, which corresponds to the peak of the disk luminosity, but our results are not qualitatively changed unless $n$ is at least an order of magnitude larger than we assume.

From the classical $\alpha$-disk model \citep{1973A&A....24..337S}, the inflow timescale of the disk is
\begin{equation}
T_{\rm inflow} = \frac{1}{\alpha} \times \left(\frac{H}{R}\right)^{-2} \times P_a,
\end{equation}
where $\alpha$ is the viscosity parameter of the disk, $H/R$ is the disk aspect ratio at a radius $R$, and $P_a$ is the orbital period at the outer edge of a disk of size $a_{\rm disk}$. Therefore, in deep plunging TDEs, because $a_{\rm disk}\ll a_{\rm mb}$, $P_a \ll P_{\rm mb}$, and the inflow timescale is likely to be shorter than the circularization / fallback timescale. In this situation the onset of accretion is therefore rapid; it can easily occur before the disk circularizes fully. 

For a TDE around a black hole with $M<~$few$~\times10^7 M_\odot$, the mass fallback rate is greater than the Eddington accretion rate for the first weeks to years. When the circularization is efficient due to the strong stream collisions as discussed above, the gas supply rate to the disk is also super-Eddington. Modern simulations of super-Eddington accretion flow \citep[e.g.,][]{2015MNRAS.447...49S} show that such disks are geometrically and optically thick, which together with strong outflows completely obscure the inner region of the disk. Studies of the outflow structure and rate of super-Eddington accretion are far from complete, so for simplicity we assume that $\sim 10\%$ of the fallback mass flows out in a wind, launched from near the innermost stable circular orbit (ISCO) at the orbital speed there (for motivation see \citealt{1982MNRAS.199..883B, 2012JPhCS.372a2040T}). At the circularization time, the wind from the power-law decay part of the fallback (starting at $\sim 1.5 P_{\rm mb}$) has reached out to $r_{\rm out}=(T_{\rm circ}~-~1.5P_{\rm mb})~\times~v_K$. Taking the time-averaged $\overline{\dot{M}}_{\rm wind}~\simeq~0.1~\times~\overline{\dot{M}}_{\rm fallback}$, the optical depth of the wind beyond $a_{\rm disk}$ is:
\begin{equation}
\tau =  \int_{a_{\rm disk}}^{r_{\rm out}} \kappa_{\rm es} \rho(r) dr =  \int_{a_{\rm disk}}^{r_{\rm out}} \kappa_{\rm es} \frac{\dot{M}_{\rm wind}}{4 \pi r^2 v_K} dr \simeq \frac{\kappa_{\rm es} \overline{\dot{M}}_{\rm fallback}}{40 \pi v_K} (\frac{1}{a_{\rm disk}}-\frac{1}{r_{\rm out}}), 
\end{equation}
where $\kappa_{\rm es}=0.34~{\rm cm}^2~{\rm g}^{-1}$ is the electron scattering opacity, and $v_K= c/2$ at the Schwarzschild ISCO.  Using $\dot{M}_{\rm fallback}(t)$ from \citet{1989ApJ...346L..13E}, and given $r_{\rm out}~\gg~a_{\rm disk}$, we find that $\tau\sim\kappa_{\rm es}\overline{\dot{M}}_{\rm fallback}/(40{\pi}{v_K}a_{\rm disk}) <1$ except in the case of the most extreme super-Eddington accretion, for which $\tau \sim 1$.  Thus the wind beyond the disk does not significantly change the observed temperature. 

We therefore compute the effective temperature $T_{\rm eff}$ of the system using a photospheric size $\sim a_{\rm disk}$:
\begin{equation}
L = 4\pi \sigma \ a^2_{\rm disk} \ T^4_{\rm eff} = \eta \dot{M}_{\rm acc} c^2 \leq L_{\rm Edd}.
\end{equation}
Here $L$ is the luminosity of the disk, $\sigma$ is the Stefan-Boltzmann constant, and we cap the luminosity at Eddington; this is consistent with some simulations \citep{2014MNRAS.441.3177M}, although other simulations find that the luminosity could be somewhat higher \citep{2014ApJ...796..106J, 2015MNRAS.447...49S}, which would imply higher temperatures. The accretion efficiency $\eta$ is always taken to be $\sim0.1$, which is comparable to the specific binding energy at the ISCO of a moderately spinning black hole. This calculation of the photosphere size is applicable to super-Eddington accretion. If the fallback rate at $5 P_{\rm mb}$ is sub-Eddington, we use the fallback rate to calculate the peak accretion power, but $T_{\rm eff} \propto \dot{M}^{1/4}$ is insensitive to $\dot{M}_{\rm acc}$.

\begin{figure}
\centering
 \includegraphics[width=6in]{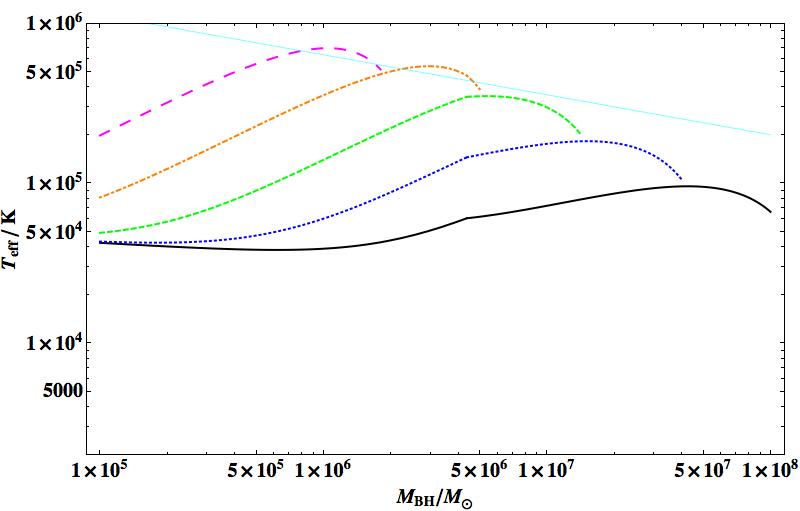}
\caption{The effective temperature of the debris disk. The color and line style scheme are the same as in Fig. 2. The light blue thin line is the temperature of a thin disk at its ISCO. This figure shows that soft X-ray temperature TDEs are produced by stars in deep plunging orbits with $\beta >3 $ (green-dashed line and lines with higher $T_{\rm eff}$).}
\label{Temperature}
\end{figure}

We plot $T_{\rm eff}$ in Figure~\ref{Temperature}. $T_{\rm eff}$ increases with increasing $\beta$.  To reach the $T\gtrsim {\rm few} \times10^5 \it $~K seen in X-ray selected TDEs, it is necessary that $\beta > 3$ and $M\lesssim 5 \times10^6 M_\odot$.  Note that disk radiative transfer effects can alter the spectrum so that the best-fit Planck temperature can be up to $\sim 2$ times larger than the effective temperature (see, Table 1 of \citealt{2005ApJ...621..372D}). Therefore, very large $\beta$ may not be required to explain TDEs with $T>5 \times 10^6$~K. 

\section{Discussion and Summary}

We consider the tidal disruption of a main-sequence star by a non-spinning SMBH, and in particular the tidal stream intersection and circularization driven by apsidal precession of the debris orbit. We show that because apsidal precession is greater for more massive black holes and deeper orbital penetration, such encounters also lead to stream intersection closer to the hole.  This strengthens shocks, which enhances energy losses and decreases the circularization timescale. The resulting super-Eddington accretion disk is small and has a short inflow timescale, so the initial disk temperature is high.

For a fixed black hole mass, stars disrupted in closer orbits produce higher temperature events. However, for fixed $\beta$ this disk temperature does not decrease monotonically with increasing black hole mass; it appears that for X-ray detected TDEs it is necessary that $\beta > 3$ and $M \lesssim 5\times10^6 M_\odot$. This is consistent with the mass distribution of the small number of X-ray observed TDE flares reported in \citet{2014arXiv1410.7772S}. Such high $\beta$ encounters are most likely to be produced when the orbital phase space of the stars is not depleted by disruptions.  In this case, the probability of an event with $\beta>\beta_0$ scales as $1/\beta_0$.  If, in contrast, the inner galactic zone is depleted, diffusion processes tend to produce low-$\beta$ encounters (e.g., \citealt{2005PhR...419...65A,2013ApJ...777..133M}). Therefore we expect that  $<1/3$ of the fully disruptive events reach a temperature high enough to produce substantial X-rays.  Reprocessing from, e.g., dust could reduce this fraction. Nonetheless, this ratio is consistent with the ratio of the observed X-ray TDE rate \citep{2002AJ....124.1308D} and optical/UV TDE rate \citep{2014ApJ...792...53V}, though both estimates bear uncertainties due to the small sample sizes and modeling assumptions. We note that not many TDEs have been observed at their peaks with X-ray telescopes, so a more complete search on X-ray TDEs will greatly improve our understanding. 

There are other models proposed to explain X-ray TDEs. For example, inferred temperature differences among TDEs may be partially due to viewing angle dependent obscuration by the disk \citep{2005PASJ...57..513W, 2014ApJ...781...82C} or wind \citep{2014MNRAS.441.3177M, 2015MNRAS.453.3213S}, leading in the extreme case to a jetted TDE when viewed down the jet \citep{2015MNRAS.454L...6M}. If the disk is geometrically thin, unobscured, and the accretion rate can stay relatively constant with radius, then the disk temperature can be dominated by the temperature near disk inner regions \citep{2014ApJ...783...23G}. Non-thermal processes such as electron scattering and Comptonization can drive part of the TDE disk to have a high temperature \citep{2002ApJ...576..753L}. Thermal or non-thermal X-ray emission could also arise from either the disk corona or jet, where a similar ambiguity exists for the origin of SgrA* radio and X-ray emission \citep{2000A&A...362..113F}. 

Around a spinning black hole, the nodal precession of the debris stream can delay substantially the time of first stream intersection \citep{2013ApJ...775L...9D,2015arXiv150105207H,2015ApJ...809..166G}, and this likely also delays the onset of accretion. As debris streams may only partially collide, circularization can take longer and the peak luminosity will be lower than it would be around a non-spinning hole. However, if the initial stellar orbit is aligned with the black hole spin, the stream intersection will be very similar to that discussed in this paper. The extra precession produced by black hole spin on a retrograde orbit can draw the intersection closer to the hole. This produces an even smaller disk and thus promotes faster onset of accretion and increases disk temperature.

LD and JCM acknowledge NASA/NSF/TCAN (NNX14AB46G), NSF/XSEDE/TACC (TGPHY120005), and NASA/Pleiades (SMD-14-5451). MCM was supported in part by NSF (AST-1333514). We are grateful for conversations with T. Alexander, R. Blandford, J. Cannizzo, B. Cenko, R. Cheng, S. Gezari, J. Guillochon, S. Komossa, J. Krolik, M. MacLeod, E. Ramirez-Ruiz, H. Shiokawa, N. Stone and S. van Velzen. We also thank the anonymous referee for helpful comments.

\clearpage

\end{document}